\documentstyle[12pt]{article}

\def\beq{\begin{equation}}
\def\eeq{\end{equation}}
\def\bea{\begin{eqnarray}}
\def\beas{\begin{eqnarray*}}
\def\beau{\begin{equation} \begin{array}{rcl}}
\def\eea{\end{eqnarray}}
\def\eeas{\end{eqnarray*}}
\def\eeau{\end{array} \end{equation}}
\def\bay{\begin{array}}
\def\eay{\end{array}}
\def\ds{\displaystyle}

\newcommand{\erre}{{\hbox{{\rm I}\kern-.2em\hbox{\rm R}}}}
\def\r4{{\mbox{\footnotesize\erre}}^{4}}

\def\half{{1 \over 2}} 
\def\Tr{ \mbox{Tr} \, }

\def\G{ {\cal G} }

\newcommand{\ra}{{\rightarrow}}

\newcommand{\sgf}{S_{\rm gf}}

\newcommand{\Bpar}{B_{/ \kern-.2em /}}

\newcommand{\parall}{{/ \kern-.2em /}}

\newcommand{\OAhat}{{\widehat{\Omega}_{A}}{}}
\newcommand{\OBhat}{{\widehat{\Omega}_{B}}{}}
\newcommand{\Oehat}{{\widehat{\Omega}_{\eta}}{}}
\newcommand{\Ophat}{{\widehat{\Omega}_{\psi}}{}}

\newcommand{\Shat}{{\widehat{\Sigma}}}

\newcommand{\Bhat}{{\widehat{B}}}

\newcommand{\dfp}[2]{ {{\delta #1 } \over {\delta #2 }} }

\newcommand{\dde}[1]{\partial_{#1}}

\newcommand{\de}{\partial}
\newcommand{\di}{{\rm d}}
\newcommand{\did}{{\rm d}^{\dagger}}
\newcommand{\demu}{\dde{\mu}}

\newcommand{\da}{{\rm d}_A}

\newcommand{\rifbibl}[1]{ #1 }
\newcommand{\ap}[3]{\rifbibl{Ann. Phys. (NY)} {\bf #1} (#2) #3}
\newcommand{\cmp}[3]{\rifbibl{Commun. Math. Phys.} {\bf #1} (#2) #3}
\newcommand{\nc}[3]{\rifbibl{Nuovo Cim.} {\bf #1} (#2) #3}
\newcommand{\np}[3]{\rifbibl{Nucl. Phys.} {\bf #1} (#2) #3}
\newcommand{\pl}[3]{\rifbibl{Phys. Lett.} {\bf #1} (#2) #3}
\newcommand{\pr}[3]{\rifbibl{Phys. Rev.} {\bf #1} (#2) #3}

\begin{document}  
  
\begin{titlepage}
\vskip 1cm  
\begin{flushright}  
{IFUM 564/FT}\\  
\end{flushright}  
\vskip .5cm  
\begin{center}  
   
{\Large \bf Renormalization Transformations \vspace{.2cm} \\
	 of the 4D BFYM Theory}\\   
    
\vskip 1cm 
{\bf Alberto Accardi}\footnote{\label{mail} 
E-mail: {\it laurqcd@astmi3.uni.mi.astro.it}}, 
$\mbox{\bf Andrea Belli}^{\ref{mail}}$  
\vskip .1cm  
{\sl Dipartimento di Fisica, Universit\`a di Milano \\  
Via Celoria 16 \ \ 20133 \ Milano \ \ ITALY}  
\end{center}  
\vskip 1cm
\abstract{We study the most general renormalization transformations for 
the first-order formulation of the Yang--Mills theory. We analyze, in 
particular, the trivial sector of the BRST cohomology of
two possible formulations of the model: the standard one and the extended one.
The latter is a promising starting point for the interpretation of the 
Yang--Mills theory as a deformation of the topological BF theory.
This work is a necessary preliminary step towards any perturbative 
calculation, and completes some recently obtained results.}  
\vskip 1cm
\centerline{PACS: 11.10, 11.15.B}
\centerline{Keywords: Yang--Mills, Topological BF Theory, 
Algebraic Renormalization}
\vfill  
\end{titlepage}  
\setcounter{section}{0}  
\section{Introduction}  
\setcounter{equation}{0}  
\addtolength{\baselineskip}{0.3\baselineskip}   
The confinement problem in QCD has been recently studied by reformulating 
the Yang--Mills action through the first-order formalism \cite{fmz}. This 
model, named ``gaussian'' BFYM, is described by the action
\beq
	S = \Tr \int_{\r4} \bigg( 
		iB\wedge F + g^2 B\wedge *B \bigg) \ ,
 \label{BFYM action}
\eeq
where $F$ is the field-strength and $B$ is a 2-form. It is easy to see that 
this action is on-shell equivalent to the classical YM action. This 
formulation \cite{fmz,ccgm} allows 
the introduction of 't Hooft-like order-disorder 
parameters \cite{thooft}, and leads to an explicit realization of the 
't Hooft picture of the vacuum as a dual superconductor, in which
the BFYM theory appears to be in the confining phase 
\cite{fmz}. \\
Another formulation of the theory is possible. Its construction relies on 
the observation that the zero coupling limit of (\ref{BFYM action}) is the 
topological pure BF theory \cite{Horowitz,BlauThompson}, whose topological 
properties are closely related to a further symmetry, named ``topological'', 
besides the gauge invariance. Then, by introducing a 1-form $\eta$, 
another first-order formulation may be defined, which is again on-shell 
equivalent to the YM theory but has a further symmetry that acts on $A$ and 
$B$ like the topological symmetry of the pure BF theory \cite{ccfmtz}:
\beq
	S = \Tr \int_{\r4} \bigg[ iB\wedge F 
		+ g^2 \Big(B-\da\eta\Big)\wedge *\Big(B-\da\eta\Big) \bigg] \ ;
 \label{BFYMeta action}
\eeq
we will call this formulation ``extended''. Then the model can be 
interpreted as a deformation of the BF topological theory \cite{ccfmtz}.

The question arises if the classical equivalence of these two models with 
the Yang-Mills extends also at the quantum level, and of what are their 
perturbative properties. A study of the 3D case was performed in 
\cite{abmz}, where the quantum equivalence was explicitly demonstrated 
together with the absence of anomalies by algebraic and power-counting 
methods, and the complete renormalization transformations were given.
The anomalies and the quantum equivalence in 4D have been treated in 
\cite{Sorellaandco} and a generalization to all dimensions can be found in
\cite{Henneaux}, where no power-counting argument is used. The equivalence 
has been studied also in \cite{mz}, where an explicit one-loop computation 
of the $\beta$-function in the gaussian formulation has been carried out 
showing that it is equal to the YM case, and in \cite{ccfmtz} with 
background field methods. \\
In this letter we wish to complete the algebraic analisys of 
\cite{Sorellaandco,Henneaux} by studying the trivial sector of the BRST 
cohomology at ghost number zero, i.e. the counterterms that can be written 
as BRST variations of some field functional. Following 
\cite{abmz} we will give the complete renormalization transformations 
needed to absorb all the invariant counterterms, and will study the  
restriction arising in the Landau gauge for both formulations. We 
will also show  that a certain class of counterterms, though algebraically 
allowed to appear, have vanishing coefficient to all orders. 
In section 2 we will study the gaussian formulation and in section 3 we 
will address the more involved extended formulation.

We will adopt the following notation. The field strength is
$F=\di A + A\wedge A$, the covariant derivative is $\da \omega =
\di \omega + [A,\omega]$. All the fields are in the adjoint representation
of the gauge group and the generators are taken to be antihermitean.
The exterior product of a p-form with a q-form is
$\omega\wedge\lambda ={1\over p!q!} \varepsilon_{\mu_1 \ldots \mu_{p+q}}
\omega_{\mu_1\ldots\mu_p}\lambda_{\mu_{p+1}\ldots\mu_{p+q}} \di x_{\mu_{1}} 
\ldots \di x_{\mu_{p+q}}$; the Hodge dual of a p-form is  
$* \omega = {1 \over {(4-p)!}} \varepsilon^{\mu_{1}\mu_{2}\mu_{3}\mu_{4}} 
\omega_{\mu_{1}\ldots\mu_{p}} \di x_{\mu_{p+1}} \ldots \di x_{\mu_{4}}$. 
Moreover $[\omega,\lambda]$ will indicate the graded commutator between the 
two forms $\omega$ and $\lambda$, and we will omit the wedge product 
between forms.
   
\section{Gaussian formulation}  
 \setcounter{equation}{0}  

The gaussian model (\ref{BFYM action}) is invariant with respect to the 
gauge symmetry
\beq
  \bay{rcl}
	\delta_g A &=& - \da \varepsilon  \\
	\delta_g B &=& - [B,\varepsilon]  \ , \\
  \eay
 \label{gauge invariance}
\eeq
where $\varepsilon$ is a local Grassmann-odd adjoint-valued zero-form.
Following the BRST quantization procedure \cite{becchi} we introduce a 
couple of ghost and antighost $(c,\bar{c})$ and 
the auxiliary field $h_A$ and we define the BRST transformation $s$:
\beq  
	\bay{rclcrcl}  
		s \, A &=& - \da c\ , & \hspace*{2cm} & 
			s \, B &=& - [B,c] \ , \\  
		s \, c &=& \half [c,c] \ , &&&& \\  
		s \, \bar{c} &=& h_A \ , &&&& \\  
		s \, h_A &=& 0 \ , &&&&
	\eay   
\eeq
which is off-shell nihilpotent.
Then we define the gauge-fixing lagrangean, choosing the covariant 
Landau gauge\footnote{The main motivation for this choice is the richer 
algebraic structure present in this gauge, and its convenience when 
performing perturbative calculations.} $\did A = 0$:   
\beq  
	\sgf =  s \ \Tr \int_{\r4} \bar{c}  * \did A \ .
\eeq 
Furthermore we introduce a set of external sources 
coupled to the nonlinear BRST transformations:
\beq
	S_{ext} = \Tr \int_{\r4} \Big( \Omega_A  * s(A) 
		+ \Omega_B  * s(B) 
		+ \Omega_c  * s(c) \Big) \ .
 \label{lagrangiana classica}  
\eeq  
Eventually, the tree-level action is  
\bea
	\Sigma \Big[ A,B,c,\bar{c},h_A,\Omega_A,\Omega_B,\Omega_c \Big] &=& 
		S_{BFYM} + S_{gf} + S_{ext} = \nonumber \\
	&=& \Tr \int_{\r4} \Big[ i B  F + g^2 B  *B    
		+ \bar{c}  * \did \da c + h_A  * \did A + \nonumber \\
	&& \hspace*{1cm} - \Omega_A * \da c 
		- \Omega_B  * [B,c] + \Omega_c  * \half\{c,c\} \Big] \ .  
 \label{classical action}  
\eea  
The dimensions, ghost-number and space-time inversion parity of 
the fields are shown in table \ref{tab: bf+b2 campi}.  \\
\begin{table}[thb]  
  \beas  
   \bay{|l||c|c|c|c|c|c|c|c|}  \hline  
	& A & B & c & \bar{c} & h_A & \Omega_A & \Omega_B & \Omega_c    
		\\ \hline\hline  
	\mbox{dimension} & 1 & 2 & 0 & 1 
		& 2 & 2 & 2 & 4   \\  \hline  
	\mbox{Ghost number} & 0 & 0 & 1 & -1 & 0 & -1 & -1 & -2  \\  \hline  
	\mbox{Space-time parity} & - & + & + & + & + & - & + & +  \\  \hline 
   \eay  
  \eeas 
 \vspace{-11.5mm}  \\ 
 \caption{dimensions, ghost-number and parity   
	of the fields  \label{tab: bf+b2 campi} }  
\end{table}  \\ 
Due to the gauge invariance (\ref{gauge invariance}), the action 
(\ref{classical action}) satisfies the Slavnov--Taylor condition:
\beq  
	{\cal S} (\Sigma) = 0  \ ,\label{ST identity}
\eeq  
where  
\beq  
	{\cal S}(\Sigma)    = \Tr \int {\rm d}^4x   
		\left( \dfp{\Sigma}{A_\mu} \dfp{\Sigma}{\Omega_{A\mu}}  
		+ \dfp{\Sigma}{B_{\mu\nu}} \dfp{\Sigma}{\Omega_{B\mu\nu}}   
		+ h_A \dfp{\Sigma}{\bar{c}}  
		+ \dfp{\Sigma}{c} \dfp{\Sigma}{\Omega_c}  \right) \ .
\eeq  
Moreover it satisfies the following constraints:
\bea 
	\dfp{\Sigma}{h_A} &=& \demu A_\mu \ ,  \label{cond1} \\
	\bar{\cal G} \Sigma &\equiv& \dfp{\Sigma}{\bar{c}} 
		+ \demu \dfp{\Sigma}{\Omega_{A\mu}} = 0  \ , \label{cond2}\\ 
	\G \Sigma &\equiv& \!\!\!\! \int \! \di^4x \! \left( \dfp{\Sigma}{c}     
		+ \left[ \bar{c} , \dfp{\Sigma}{h_A} \right] \right) \! = 
		\! \int \!
		\di^4x \bigg( \! \Big[ {\Omega_A}_\mu , A_\mu \Big] + \Big[ 
		{\Omega_B}_{\mu\nu} , B_{\mu\nu} \Big] 
		- \Big[ \Omega_c , c \Big] \! \bigg) , \label{cond3}\\
	W^{rig}\Sigma &\equiv& \int \di^4x \sum_\varphi   
		\left[ \varphi,\dfp{\Sigma}{\varphi} \right]  = 0 
		\label{cond4} \hspace*{2.5cm} \varphi =\mbox{all fields} \ .
\eea
The constraint (\ref{cond2}) is obtained by commuting (\ref{cond1}) with 
the S.T. identity (\ref{ST identity}), and the constraint (\ref{cond4})
is obtained by commuting (\ref{cond3}), 
which is a peculiarity of the Landau gauge \cite{Landau gauge},  
with (\ref{ST identity}).
As a consequence of (\ref{cond2}) the antighost enter in the  
action only through the combination:  
\beq  
	\widehat{\Omega}_{A\mu} = \Omega_{A\mu} + \demu \bar{c} \ .  
 \label{Ohat}  
\eeq  
We can define the reduced action 
\beq  
	\widehat{\Sigma}[A,B,c,{\OAhat},\Omega_B,\Omega_c]  
		= \Sigma [A,B,c,\bar{c},h_A,\Omega_A,\Omega_B,\Omega_c]   
		- \int \di^4x h_A \demu A_\mu \ ,
\eeq
so that with respect to these new variables, the constraints 
(\ref{cond1}-\ref{cond4}) become
\bea 
	\dfp{\Shat}{h_A} &=& 0  \label{cond1p} \ , \\
	\dfp{\Shat}{\bar{c}} &=& 0 \label{cond2p} \ , \\ 
	\int \dfp{\Shat}{c} &=& \int\di^4x \bigg( \Big[    
		{\OAhat}_\mu , A_\mu \Big] + \Big[ {\Omega_B}_{\mu\nu} , 
		B_{\mu\nu} 
		\Big] - \Big[ \Omega_c , c \Big] \bigg) \label{cond3p} \ , \\
	W^{rig}\Shat &=& \int \di^4x \sum_\varphi   
		\left[ \varphi,\dfp{\Shat}{\varphi} \right]  = 0 
		\label{cond4p} \hspace*{2.5cm} \varphi =\mbox{all fields} \ .
\eea
By introducing the linearized S.T. operator
\beq  
	\Bhat_\Shat = \Tr \!\! \int \! \di^4x \!   
		\left( \dfp{\Shat}{A_\mu} \dfp{}{\OAhat_\mu}  
		+ \dfp{\Shat}{\OAhat_\mu} \dfp{}{A_\mu}  
		+ \dfp{\Shat}{B_{\mu\nu}} \dfp{}{{\Omega_B}_{\mu\nu}}  
		+ \dfp{\Shat}{{\Omega_{B\mu}}} \dfp{}{B_\mu}  
		+ \dfp{\Shat}{c} \dfp{ }{{\Omega_c}}  
		+ \dfp{\Shat}{{\Omega_c}} \dfp{}{c}  \right) , 
 \label{ST operator}
\eeq  
the S.T. identity (\ref{ST identity}) is rewritten as 
\beq 
	{\cal S}(\Sigma) = \half \Bhat_\Shat \Shat = 0 \ ,
\eeq
which, in turn, implies the nihilpotency of the S.T. operator:
\beq
	\Bhat_\Shat \Bhat_\Shat = 0 \ .
\eeq

As it is well known \cite{PiguetSorella95} the anomalies and 
the counterterms are described by the local
cohomology of the S.T. operator {\it modulo} total derivatives. In fact,
both the possible breaking $\Delta$ 
of the S.T. identity and the invariant counterterms 
have to satisfy the consistency condition
\beq
	\Bhat_\Shat \Delta + \di \widetilde{\Delta} = 0
\eeq
where $\Delta$ is a local 4-form of ghost number $1$ and $0$ respectively, and 
dimension 4 thank to the QAP \cite{qap}, and $\widetilde{\Delta}$ is a
3--form of ghost number $2$ and dimension $3$. The solution of these two 
cohomology problems have been worked out in \cite{Sorellaandco,Henneaux},
whose result is that the cohomology of the S.T. operator 
(\ref{ST operator}) is isomorphic to the YM case. Then the only anomaly 
allowed by the theory is the ABBJ one \cite{ABBJ}, which is absent in our 
case since all the fields are in the adjoint representation. Moreover, the 
only physical renormalization is the coupling constant one, the related 
counterterm being $\Delta = \frac{1}{g^2} F*F$ which is 
$(\Bhat_\Shat \mbox{ {\it modulo} } \di)$--equivalent to a multiple of
$g^2 B *B$.

It remains to analyze the trivial counterterms; they are of the form
\beq
	\Bhat_\Shat \ \Tr \int_{\r4} \widetilde{\Delta} \ ,
\eeq
where $\widetilde{\Delta}$ is a local 4-form of ghost number $-1$ and 
dimension 4. Let us introduce the following  notation:
\beq
	N_\varphi = \int_{\r4} \varphi * \dfp{}{\varphi} \ \ \  ; \ \ \
		N_{\varphi \rightarrow \omega} = \int_{\r4} \omega * 
		\dfp{}{\varphi} \ .
\eeq
Then the trivial counterterms can be expressed as
\beq
  \bay{l}  
    \Bhat_\Shat ( \Tr \int \OAhat*A) = (N_A - N_{\OAhat}) \Shat 
    \equiv {\cal N}_A \Shat \ ,\\  
    \Bhat_\Shat ( \Tr \int \Omega_B*B) = (N_B - N_{\Omega_B}) \Shat 
    \equiv {\cal N}_B \Shat \ , \\  
    \Bhat_\Shat ( \Tr \int \Omega_B\di A) = (N_{B\rightarrow *\di A} 
	+ N_{\OAhat \rightarrow *\di \Omega_B} ) \Shat 
	\equiv {\cal N}_{rot}^{(1)}  \Shat  \ , \\  
    \Bhat_\Shat ( \Tr \int \Omega_B[A,A]) = (N_{B\rightarrow *[A,A]} 
	+ 2 N_{\OAhat \rightarrow *[A,\Omega_B]} ) \Shat
	\equiv {\cal N}_{rot}^{(2)}  \Shat  \ . \\  
    \Bhat_\Shat ( \Tr \int \Omega_BB) = (N_{B\ra*B} - 
	N_{\Omega_B \ra * \Omega_B}) 
	\Shat \equiv {\cal N}^*_B \Shat \ , \\  
    \Bhat_\Shat ( \Tr \int \Omega_B*\di A) = (N_{B\rightarrow \di A} 
	- N_{\OAhat \rightarrow \di \Omega_B} ) \Shat 
	\equiv {\cal N}_{rot}^{*(1)}  \Shat  \ , \\  
    \Bhat_\Shat ( \Tr \int \Omega_B*[A,A]) = (N_{B\rightarrow [A,A]} 
	- 2 N_{\OAhat \rightarrow [A,\Omega_B]} ) \Shat
	\equiv {\cal N}_{rot}^{*(2)}  \Shat  \ . \\  
    \Bhat_\Shat ( \Tr \int \Omega_c*c) = (- N_c + N_{\Omega_c}) \Shat 
    	\equiv{\cal N}_c \Shat  \ , \\  
  \eay
 \label{controt con N}  
\eeq
The ${\cal N}_c \Shat$ counterterm is furthermore excluded by the ghost 
equation (\ref{cond3p}). Then all the counterterms can be absorbed at the
order $\hbar^n$ by the 
following renormalization transformation, assuming that the renormalization
process has been carried out till the order $\hbar^{n-1}$ 
(details can be found in
\cite{abmz}, which deals with the 3D case):
\beq  
  \bay{rcl}
	A &=& A_R + \hbar^n z_A A_R  \ , \\
	B &=& B_R + \hbar^n z_B B_R + \hbar^n z_{BA} *\di A_R
		+ \hbar^n z_{BAA} * \frac{1}{2}[A_R,A_R] + \\
	&& + \hbar^n z^*_B * B_R
		+ \hbar^n z^*_{BA} \di A_R
		+ \hbar^n z^*_{BAA} \frac{1}{2}[A_R,A_R]
		\ , \\
	c &=& c_R \ , \\
	\OAhat &=& \OAhat_R - \hbar^n z_A {\OAhat}_R 
		+ \hbar^n z_{BA} * \di {\Omega_B}_R 
		+ \hbar^n z_{BAA} * [A_R,{\Omega_B}_R] + \\
	&& - \hbar^n z^*_{BA} \di {\Omega_B}_R 
		- \hbar^n z^*_{BAA} [A_R,{\Omega_B}_R] 
		\ , \\
	{\Omega_B} &=& {\Omega_B}_R  - \hbar^n z_B {\Omega_B}_R
		- \hbar^n z^*_B {\Omega_B}_R \ , \\
	{\Omega_c} &=& {\Omega_c}_R \ , \\ 
	g &=& g_R + \hbar^n z_g g_R \ . \\
  \eay
\eeq
We note that the terms containing $F$ get contributions not only from 
the rescaling 
of $g$ and $A$, as it happens in the YM case, but also from the rotation of 
the $B$-field, that gives {\it a priori} two different weights to $\di A$
and $[A,A]$.

Let us comment about the terms of coefficient $z^*$. These terms produce 
counterterms of the type $\epsilon_{\mu\nu\rho\sigma} B_{\mu\nu}
B_{\rho\sigma}$, $B_{\mu \nu} \demu A_\nu$ and 
$B_{\mu \nu} [A_\mu,A_\nu]$ which alter the tensorial structure of the 
1PI functions $\Gamma_{BB}$ $\Gamma_{BA}$ and $\Gamma_{BAA}$ with 
respect to the corresponding Feynman 
rules. But it is possible to show that starting from the Feynman 
rules stemming from the action (\ref{classical action}) these structure are 
not generated at any order of perturbation, meaning that all the 
renormalization constants $z^*$ are equal to zero to all orders.
Indeed if we consider, for example, any given diagram of the two point 
function $\Gamma_{BA}$ we see that the sum of the number of  vertices $BAA$ 
and the number of propagators $BA$, that are the 
only Feynman rules proportional to the totally antisymmetric tensor 
$\epsilon_{\mu\nu\rho\sigma}$, is always an odd number so that the net 
result is that every diagram is proportional to this tensor and the structures 
coming from the $z^*$ terms do not appear.  \\
The situation is as follows. To obtain the equivalence with the Yang-Mills 
theory we could have equally well started from the first-order lagrangean
$iB*F + g^2 B*B$, whose $g\ra 0$ limit is again symmetric with 
respect to the topological-like symmetry $B\ra B + *\da \xi$ allowing for 
the extended formulation that will be discussed in the next section. The 
algebraic structure of each model allows for the generation of the 
counterterms corresponding to the other theory, but its Feynman rules 
do not generate them: the two theories, then, do not mix. They would do if 
we would have started from a Lagrangian of the type
$iB F \pm iB *F + g^2 B *B$, which is equivalent on shell to 
the Yang--Mills in the self-dual (anti self-dual) gauge, and whose $g\ra 0$
limit is again symmetric with respect to $B \ra B + \da \xi \mp *\da \xi$.

\section{Extended formulation}  
 \setcounter{equation}{0}  

The analysis of this formulation proceeds in the same way as the former one, 
so that we will shorten the discussions; details can be found in 
\cite{Sorellaandco}.

The action (\ref{BFYMeta action}) is invariant with respect to the 
gauge symmetry $\delta_g$, the topological symmetry $\delta_t$ and the 
further local simmetry $\delta'$ \cite{ccfmtz,Sorellaandco}
\beq\bay{rclcrclcrcl}
	\delta_g A &=&  - \da \varepsilon \ , &\hspace*{1cm}& \delta_t A &=& 0 
		\ , &\hspace*{1cm}& \delta^\prime A &=& 0 \ , \\
	\delta_g B &=& - [B,\varepsilon] \ , && \delta_t B &=& \da \theta \ , 
		&& \delta^\prime B &=& [F,\sigma] \ , \\
	\delta_g \eta &=& - [\eta,\varepsilon] \ , && \delta_t \eta &=& \theta 
		\ , && \delta^\prime \eta &=& \da \sigma \ ,
\eay\eeq
where $\varepsilon$,$\sigma$ are local 0-forms and $\theta$ is a local 
1-form.
We remark that the topological symmetry $\delta_t$ is reducible, requiring 
a second ghost generation in order to correctly quantize the theory and that
$\delta'$ is not independent of the topological symmetry as it can be seen 
by choosing $\theta = \da \sigma$. 

The BRST quantization requires for each classical symmetry 
the introduction of a couple of ghost and antighost 
and a Lagrange multiplier, called respectively 
$(c, \bar c, h_A)$, $(\psi, \bar\psi, h_B)$ and $(\rho, \bar\rho, h_\eta)$.
We have also to introduce the second generation ghosts and multiplier
$(\phi, \bar\phi, h_\psi)$ and also the pair of fields $(u, h_{\bar\psi})$
to fix a further degeneracy related to $\bar\psi$. Their 
dimension, ghost number and parity are shown in table 
\ref{extended fields}. 
Then, we define the BRST transformation
\beq\bay{rclcrclcrcl} 
	sA &=& -\da c \ , &\hspace*{.14cm}& sB &=& - [B,c] + \da \psi + [F,\rho] 
		\ , &\hspace*{.14cm}& s\eta &=& -[\eta,c] + \psi + \da \rho \ , \\
	sc &=& {1 \over 2}[c,c] \ , &&  s \psi &=& [\psi,c] + \da \phi \ , &&
		s \rho &=& [\rho,c] - \phi \ , \\
	s \bar c &=& h_A \ , && s \bar \psi &=& h_B \ , && s \bar\rho &=& h_\eta 
		\ , \\
	s h_A &=& 0 \ , && s h_B &=& 0 \ , && s h_\eta &=& 0 \ , \\
	&&&& s \phi &=& - [\phi,c] \ , &&&& \\
	&&&& s \bar\phi &=& h_\psi\ , &&&& \\
	&&&& s h_\psi &=& 0 \ , &&&& \\
	&&&& s u &=& h_{\bar\psi} \ , &&&& \\
	&&&& s h_{\bar\psi} &=& 0 &&&& 
 \label{extended BRST}
\eay\eeq
\begin{table}[tbh]
  \beas
    \bay{|l||c|c|c|c|c|c|c|c|c|c|c|c|c|c|c|c|c|} \hline  
	& A & B & \eta & c & \bar c & \psi & \bar\psi 
		& h_A & h_B & \phi & \bar\phi & h_\psi & \rho & \bar\rho 
		& h_\eta & u & h_{\bar\psi}\\ \hline\hline 
	\mbox{dimension} & 1 & 2 & 1 & 0 & 2 & 1 & 1 & 2 & 1 
		& 0 & 2 & 2 & 0 & 2 & 2 & 2 & 2  \\ \hline
	\mbox{ghost no.} & 0 & 0 & 0 & 1 & -1 & 1 & -1 & 0 & 0 
		& 2 & 2 & -1 & 1 & -1 & 0 & 0 & 1   \\ \hline
	\mbox{parity} & - & + & - & + & + & - & - & + & - 
		& + & + & + & + & + & + & + & +  \\ \hline 
    \eay
  \eeas
\vskip -.3cm
  \beas
    \bay{|l||c|c|c|c|c|c|c|} \hline 
	& \OAhat & \OBhat & \Oehat & \Omega_c & \Ophat & \Omega_\rho
		& \Omega_\phi  \\ \hline\hline 
	\mbox{dimension} & 3 & 2 & 3 & 4 & 3 & 4 & 4  \\ \hline 
	\mbox{ghost no.} & -1 & -1 & -1 & -2 & -2 & -2 & -3  \\ \hline
	\mbox{parity}  & - & + & - & + & - & + & + \\ \hline 
    \eay
  \eeas
 \vspace{-11.5mm}  \\ 
 \caption{dimensions, ghost-number and parity   
	of the fields  \label{extended fields} }  
\end{table}
which is off-shell nihilpotent:
\beq
	s^2 = 0 \ .
\eeq
Then we choose the gauge-fixing conditions:
\beq
	\did A = 0, \ \ \ \ \did B = 0, \ \ \ \ \did \eta = 0 \ ,
\eeq
and define the gauge-fixing lagrangean in the Landau gauge:
\beq
	S_{gf} = s \ \Tr \int_{\r4} \left[ \bar c * \did A
		+ \bar\psi * \did B +\bar\rho * \did \eta
		+ \did \bar \psi * u
		+\bar\phi * \did \psi \right] \ .
\eeq
Finally, we introduce the external sources $\Omega$ coupled to the 
nonlinear BRST variations:
\bea
	S_{ext}&=& \! \Tr \int_{\r4} \bigg( -{\Omega_A} * \da c
		+ {1\over2} {\Omega_B} * (\da \psi - [B,c] + [F,\rho] )
		+ {\Omega_\eta} * (\psi - [\eta,c] + \da \rho) + \nonumber \\
	&& \hspace*{1cm} 
		+ {\Omega_\psi} * (\da\phi + [\psi,c]) + 
		{\Omega_\rho} * (-\phi + [\rho,c]) 
		+ {1\over2}{\Omega_c} * [c,c]
		-{\Omega_\phi} * [\phi,c] \bigg). 
\eea
The complete action
\beq
	\Sigma = S_{BFYM\eta} + S_{gf} + S_{ext}
\eeq
satisfies the Slavnov-Taylor identity
\beq
	{\cal S}(\Sigma)=0,
 \label{ext ST}
\eeq
where
\bea
	{\cal S}(\Sigma) &=& \Tr \int d^4x
		\bigg({\dfp{\Sigma}{A_\mu}}{\dfp{\Sigma}{\Omega_{A\mu}}}
		+ {1\over 2}{\dfp{\Sigma}{B_{\mu\nu}}}
		{\dfp{\Sigma}{{\Omega_{B\mu\nu}}}}  
		+ {\dfp{\Sigma}{\eta_\mu}}{\dfp{\Sigma}{{\Omega_{\eta\mu}}}}
		+ {\dfp{\Sigma}{\psi_\mu}}{\dfp{\Sigma}{{\Omega_{\psi\mu}}}}
		+ {\dfp{\Sigma}{\rho}}{\dfp{\Sigma}{{\Omega_\rho}}} 
		+ \nonumber  \\
	&& \hspace*{.8cm} + {\dfp{\Sigma}{ c}}{\dfp{\Sigma}{ {\Omega_c}}}
		+ {\dfp{\Sigma}{\phi}}{\dfp{\Sigma}{{\Omega_\phi}}}
		+ h_A{\dfp{\Sigma}{\bar c}} 
		+ h_{B\mu}{\dfp{\Sigma}{\bar\psi_\mu}}
		+ h_{\bar\psi}{\dfp{\Sigma}{u}}
		+ h_\psi{\dfp{\Sigma}{\bar\phi}}
		+ h_\eta{\dfp{\Sigma}{\bar\rho}}\bigg) ,
\eea
and the following constraints \cite{Sorellaandco}:
\begin{itemize}
\item the gauge fixing conditions
\beq
  \bay{ll}
	\ds {\dfp{\Sigma}{h_A}}=\demu A_\mu \ , 
	& \ds {\dfp{\Sigma}{h_\psi}}=\demu\psi_\mu \ , \\
	\ds {\dfp{\Sigma}{h_{B\nu}}}=\demu B_{\mu\nu}-\de_\nu u \ , 
	& \ds {\dfp{\Sigma}{h_{\bar\psi}}}=\demu\bar\psi_\mu \ ,  \\
	\ds {\dfp{\Sigma}{h_\eta}}=\demu\eta_\mu \ ,
	& \ds {\dfp{\Sigma}{ u}}=\demu h_{B\mu} \ ; 
  \eay
 \label{gf cond}
\eeq
\item the antighost equations
\beq
  \bay{ll}
	\ds {\dfp{\Sigma}{\bar c}}+ \demu{\dfp{\Sigma}{{\Omega_{A\mu}}}}=0 \ , 
		& \ds {\dfp{\Sigma}{\bar\psi_\nu}} 
		+ \demu{\dfp{\Sigma}{{\Omega_{B\mu\nu}}}}
		=\de_\nu h_{\bar\psi} \ , \\
	\ds {\dfp{\Sigma}{\bar\rho}}+ \demu{\dfp{\Sigma}{{\Omega_{\eta\mu}}}}=0
		\ , & \ds {\dfp{\Sigma}{\bar\phi}} 
		- \demu{\dfp{\Sigma}{{\Omega_{\psi\mu}}}}=0 \ ;
  \eay
 \label{ag cond}
\eeq
\item the ghost equations 
\bea
	&& \hspace*{-1.5cm} \int \di^4x \bigg( {\dfp{\Sigma}{c}} 
		+\Big[{\bar c,{\dfp{\Sigma}{h_A}}}\Big]
		+ \Big[{\bar\psi_\nu,{\dfp{\Sigma}{h_{B\nu}}}}\Big]
		+ \Big[{u,{\dfp{\Sigma}{h_{\bar\psi}}}}\Big] 
		+ \Big[{\bar\rho,{\dfp{\Sigma}{h_\eta}}}\Big]
		+ \Big[{\bar\phi,{\dfp{\Sigma}{h_\psi}}}\Big]\bigg)
		=\Delta_{cl}^c \ ,  
	\label{gheq1} \\
	&& \hspace*{-1.5cm} \int \di^4x \bigg({{{\delta\Sigma}\over{\delta\phi}} 
		- \Big[{\bar\phi ,{{\delta\Sigma}
		\over{\delta h_A}}}\Big]}\bigg)=\Delta_{cl}^\phi \ ,
	\label{gheq2} \\
	&& \hspace*{-1.5cm} \int d^4x \bigg({{\delta\Sigma}\over{\delta\rho}}
		+ \Big[{A_\mu,{{\delta\Sigma}\over{\delta\psi_\mu}}}\Big]
		+ \Big[{c, {{\delta\Sigma}\over{\delta\phi}}}\Big]
		- \Big[{{\Omega_{\psi\mu}},{{\delta\Sigma}
		\over{\delta {\Omega_{A\mu}}}}}\Big] + 
		\Big[{{\Omega_\phi},{{\delta\Sigma} 
		\over{\delta {\Omega_c}}}}\Big] + \nonumber \\
	&& \hspace*{-.2cm} + 
		\Big[{\bar\phi,{{\delta\Sigma}\over{\delta\bar c}}}\Big]
		- \Big[{h_\psi,{{\delta\Sigma}\over{\delta h_A}}}\Big]\bigg)
		= 0 \ ; \label{gheq3}  		
\eea
where the classical breaking $\Delta$ are
\bea
	\Delta_{cl}^c &=& \int d^4x \bigg(-[{\Omega_{A\mu}},A_\mu]
		+ {1\over2}[{\Omega_{B\mu\nu}},B_{\mu\nu}]
		+ [{\Omega_{\eta\mu}},\eta_\mu]-[{\Omega_{\psi\mu}},\psi_\mu] 
		- [{\Omega_\rho},\rho] + \nonumber \\
	&& \hspace*{1.3cm} + [{\Omega_\phi},\phi] - [{\Omega_c},c]  \bigg) 
		\ , \\
	\Delta_{cl}^\phi &=& \int \di^4x  
		\bigg({[{\Omega_{\psi\mu}},A_\mu]
		- {\Omega_\rho}+[{\Omega_\phi},c]}\bigg) \ ;
\eea 
since these breakings are linear in the quantum fields, they do not get 
radiative corrections; 
\item the rigid gauge invariance
\beq
	W^{rig}\Sigma = \int \di^4x \sum_\varphi   
		\left[ \varphi,\dfp{\Sigma}{\varphi} \right]  = 0 
		\hspace*{2.5cm} \varphi = \mbox{all fields} \ .
\eeq
\end{itemize}
Thanks to the antighost equations (\ref{ag cond}) and to the gauge-fixing 
conditions (\ref{gf cond}) we can redefine some sources:
\beq
  \bay{lcl}
	\OAhat_\mu={\Omega_A}_\mu+\demu\bar c \ , & \hspace*{1cm} & 
	\Oehat_\mu={\Omega_\eta}_\mu+\demu\bar\rho \ , \\
	\OBhat_{\mu\nu}={\Omega_B}_{\mu\nu}+\de_{[\mu}\bar\psi_{\nu]}
		\ , & & 
	\Ophat_\mu={\Omega_\psi}_\mu-\demu\bar\phi \ ,
  \eay
\eeq
and introduce the reduced action $\Shat$:
\bea
	&& \Shat [A,B,\eta,c,\psi,\rho,\phi,{\OAhat},\OBhat,\Oehat,
		\Omega_c,\Ophat,\Omega_\rho,\Omega_\phi] = 
		\nonumber \\  
	&& \hspace*{1cm} = \Sigma [A,B,c,\bar{c},h_A,\psi,{\bar\psi},h_B,
		\rho,\bar\rho,h_\eta,\phi,{\bar\phi},h_\psi,u,h_{\bar\psi},
		\Omega_A,\Omega_B,\Omega_\eta,\Omega_c,\Omega_\psi,
		\Omega_\rho,\Omega_\phi] +  \nonumber  \\
	&& \hspace*{1.4cm} - \int \! \bigg( h_A \demu A_\mu  
		+ {h_{B[\mu}} \de_{\nu]} B_{\mu\nu} + h_\eta \demu \eta_\mu
		+ h_\psi \demu \psi_\mu 
		+ h_{\bar\psi}  \demu \bar\psi_\mu
		+ u h_B \bigg) .
\eea
The S.T. equation (\ref{ext ST}) becomes
\beq
	\Bhat_\Shat \Shat = 0 \ ,
\eeq
where
\bea
	\Bhat_\Shat &=& \Tr \int \di^4x 
		\bigg({{\delta\Shat}\over{\delta A_\mu}}
		  {\delta\over{\delta\OAhat_\mu}} 
		+ {{\delta\Shat}\over
		  {\delta\OAhat_\mu}}{\delta\over A_\mu} 
		+ {1\over 2}{{\delta\Shat}\over{\delta B_{\mu\nu}}}
		  {\delta\over{\delta\OBhat_{\mu\nu}}} 
		+ {1\over 2} {{\delta\Shat}
		  \over{\delta\OBhat_{\mu\nu}}}
		  {\delta\over{\delta B_{\mu\nu}}}
		+ {{\delta\Shat}\over{\delta\eta_\mu}}
		  {\delta\over{\delta\Oehat_\mu}} + \nonumber  \\
	&& \hspace*{1.5cm} + {{\delta\Shat}\over{\delta\Oehat_\mu}}
		  {\delta\over{\delta\eta_\mu}}
		+ {{\delta\Shat}\over{\delta\psi_\mu}}
		  {\delta\over{\delta\Ophat_\mu}}
		+ {{\delta\Shat}\over{\delta\Ophat_\mu}}
		  {\delta\over{\delta\psi_\mu}}
		+ {{\delta\Shat}\over{\delta\rho}}
		  {\delta\over{\delta{\Omega_\rho}}}
		+ {{\delta\Shat}\over{\delta{\Omega_\rho}}}
		  {\delta\over{\delta\rho}} + \nonumber \\
	&& \hspace*{1.5cm} + {{\delta\Shat}\over{\delta\phi}}
		{\delta\over{\delta{\Omega_\phi}}}
		+ {{\delta\Shat}\over{\delta{\Omega_\phi}}}
		  {\delta\over{\delta\phi}}
		+ {{\delta\Shat}\over{\delta c}}
		  {\delta\over{\delta {\Omega_c}}}
		+ {{\delta\Shat}\over{\delta {\Omega_c}}}
		  {\delta\over{\delta c}}\bigg)
\eea
and
\beq
	\Bhat_\Shat \Bhat_\Shat = 0 \ .
\eeq

The cohomology of $\Bhat_\Shat$  is once again isomorphic to 
the YM case \cite{Sorellaandco,Henneaux}, so that also this 
formulation of the BFYM is equivalent at 
quantum level to the Yang-Mills. Once again, the only anomaly term is 
the ABBJ one (equal to zero in our case) and the only physical 
renormalization is the coupling constant one through the ${1 \over {g^2}}F *F$ 
non trivial counterterm, which is equivalent to a multiple of
$g^2 \left( B - \da \eta \right) * \left( B - \da \eta \right)$.  \\
The analysis of the trivial counterterms in this case is more involved 
with respect to the gaussian formulation; indeed we have found that 
the counterterms allowed by the QAP are the 
BRST variations of the traces of the integrals of the following monomials, 
that are all parity-invariant:
\beq
    \bay{lllllll}
	a_1 \ \OAhat *A & \hspace*{1cm} & b_1 \ \OBhat *B & \hspace*{1cm} & 
		e_1 \ \Omega_c *c & \hspace*{1cm} & m_1 \ \Ophat *\psi  \\
	a_2 \ \OAhat * \eta & & b_2 \ \OBhat \di A & & e_2 \ \Omega_c *\rho & & 
		m_2 \ \Ophat * \di c \\
	& & b_3 \ \OBhat \half [A,A] & & & & m_3 \ \Ophat *[A,c]   \\
	d_1 \ \Oehat *\eta & & b_4 \ \OBhat *\di\eta & & f_1 \ \Omega_\phi * 
		\phi & & m_4 \ \Ophat *[\eta ,c]  \\
	d_2 \ \Oehat * A & & b_5 \ \OBhat *[A,\eta] & & f_2 \ \Omega_\phi *[c,c]
		& & m_5 \ \Ophat *\di\rho  \\
	& & b_6 \ \OBhat [\eta,\eta] & & f_3 \ \Omega_\phi *[c,\rho] 
		& & m_6 \ \Ophat * [A,\rho] \\
	& & b_7 \ \OBhat B & & f_4 \ \Omega_\phi * [\rho,\rho] 
		& & m_7 \ \Ophat * [\eta,\rho]  \\
	& & b_8 \ \OBhat * \di A & & & &  \\
	& & b_9 \ \OBhat * \half [A,A] & & l_1 \ \Omega_\rho * \rho & & \\
	& & b_{10} \ \OBhat \di\eta & & l_2 \ \Omega_\rho * c & & \\
	& & b_{11} \ \OBhat [A,\eta] & & & & \\
	& & b_{12} \ \OBhat * [\eta,\eta] \ . & & & &  \\
    \eay  
\eeq
As in the gaussian formulation, we can extract a subset of counterterms 
that will get zero coefficient to all orders of perturbation by inspecting 
the tensorial structure of the 1PI functions. Indeed we can easily 
demonstrate that the number of $\epsilon_{\mu\nu\rho\sigma}$ tensors in any 
1PI function with $E_A$ external legs $A$, $E_B$ legs $B$ and
$E_\eta$ legs $\eta$ is equal {\it modulo} 2 to
$E_B + E_\eta$. Therefore only the graphs with an odd number of legs 
$B$ or $\eta$ will be proportional to  $\epsilon_{\mu\nu\rho\sigma}$.
Therefore, as explained at the end of section 2, we get rid of a number of 
counterterms:
\beq
  \bay{l}
	b_7 = b_8 = b_9 = b_{10} = b_{11} = b_{12} = 0  \\
	a_2 = d_2 = m_4 = m_7 = 0 
  \eay
\eeq
at any order of perturbation. Another consequence is that the 
tensorial structure of the propagators and of the vertices present at the 
classical level is not changed by radiative corrections. 
Furthermore the ghost equations (\ref{gheq1}--\ref{gheq3}) exclude some 
of the remaining monomials requiring
\beq
	e_1 = e_2 = f_3 = f_4 = 0
\eeq
and impose the following conditions on the coefficients:
\beq
  \left\{ \bay{l}
	f_2 = - l_2 = m_3 \ , \\
	f_1 = - l_1 \ , \\
	a_1 + f_1 + m_1 - m_6 = 0 \ ;	
  \eay \right.
\eeq
hence the number of free parameters to be fixed by renormalization 
conditions is 14 (13 wave-function renormalizations and the coupling 
constant renormalization). By expressing the BRST variations of the 
remaining monomials in the same way as in (\ref{controt con N}) we can 
write at once the renormalization transformations, where the fields are
understood to be renormalized till the order $\hbar^{n-1}$:
\beq      
  \bay{rcl}
	A &=& A_R + a_1 \hbar^n A_R \ , \\
	B &=& B_R + b_1 \hbar^n B_R + b_2 \hbar^n *\di A_R
		+ b_3 \hbar^n * \frac{1}{2}[A_R,A_R] + b_4 \hbar^n *
		\di\eta_R + \nonumber \\
	&& + b_5 \hbar^n * [A_R,\eta_R] + b_6 \hbar^n * [\eta_R,\eta_R] \ , \\
	\eta &=& \eta_R + d_1 \hbar^n \eta_R \ , \\
	c &=& c_R \ , \\
	\psi &=& \psi_R - m_1 \hbar^n \psi_R - m_2 \hbar^n \di c_R - f_2
		\hbar^n [A_R, c_R] - m_5 \hbar^n \di\rho_R + \nonumber \\
	&& - (a_1 + f_1 + m_1) \hbar^n [A_R,\rho_R] \ , \\
	\rho &=& \rho_R + f_1 \hbar^n \rho_R + f_2 \hbar^n c_R \ , \\
	\phi &=& \phi_R + f_1 \hbar^n \phi_R + f_2 \hbar^n [c_R,c_R] \ , \\
	\OAhat &=& \OAhat_R - a_1 \hbar^n {\OAhat}_R 
		- b_2 \hbar^n * \di \OBhat_R + b_3 \hbar^n * [A_R,\OBhat_R] 
		+ b_5 \hbar^n * [\eta_R, \OBhat_R] + \nonumber \\
	&& - f_2 \hbar^n [c_R,\Ophat_R] - (a_1 + f_1 + m_1) \hbar^n 
		[\rho_R,\Ophat_R] \ , \\
	\OBhat &=& \OBhat_R - b_1 \hbar^n \OBhat_R \ ,\\
	\Oehat &=& \Oehat_R - d_1 \hbar^n \Oehat_R - b_4 \hbar^n * \di \OBhat_R 
		+ b_5 \hbar^n * [A_R,\OBhat_R] 
		+ 2 b_6 \hbar^n * [\eta_R, \OBhat_R] \ , \\
	\Omega_c &=& \Omega_{cR} - f_2 \hbar^n \Omega_{\rho R} + 2 f_2 
		\hbar^n [c,\Omega_{\phi R}] - m_2 \hbar^n \di\Ophat_R + f_2
		\hbar^n *[A_R,*\Ophat_R] \ , \\
	\Ophat &=& \Ophat_R + m_1 \hbar^n \Ophat_R \ , \\
	\Omega_\rho &=& \Omega_{\rho R} - f_1 \hbar^n \Omega_{\rho R} - m_5 
		\hbar^n \di\Ophat_R + (a_1 + f_1 + m_1) \hbar^n * 
		[A_R, *\Ophat_R] \ , \\
	\Omega_\phi &=& \Omega_{\phi R} - f_1 \hbar^n \Omega_{\phi R} \ , \\
	g &=& g_R + \hbar^n z_g g_R \ .
  \eay   
\eeq

\section{Conclusions}  

We have analyzed the trivial counterterms of both the gaussian and 
the extended 
formulations of the BFYM theory in four dimensions, giving the full 
structure of the wave-function renormalizations and exploiting the 
restrictions arising in the Landau gauge. We have also found some 
restrictions to the tensorial structure of the 1PI functions in both 
formulations due to our choice of the classical Lagrangean, 
and consequently we have found a subclass of algebraically 
allowed counterterms that have nonetheless coefficients equal to zero 
to all orders of perturbation.  \\
This ends the algebraic analysis of the perturbative renormalization 
of the theory, completing the results of \cite{Sorellaandco,Henneaux} 
who studied the anomaly and the physical parameters renormalization.

\vskip 1.5cm  
\leftline{\bf\large Acknowledgments}  
\vskip .5cm  
The authors are grateful to M. Martellini and M. Zeni for many 
useful suggestions and discussions.
   
\newpage  
  

\end{document}